# A Strategy for Advancing Research and Impact in New Computing Paradigms

Rajkumar Buyya[1], Sukhpal Singh Gill[2], Satish Narayana Srirama[3,4], Rami Bahsoon[5], and San Murugesan[6]

[1]Cloud Computing and Distributed Systems (CLOUDS) Laboratory, School of Computing and Information Systems, The University of Melbourne, Australia
[2]School of Electronic Engineering and Computer Science, Queen Mary University of London, UK
[3]School of Computer and Information Sciences, University of Hyderabad, India.
[4]Mobile & Cloud Lab, Institute of Computer Science, University of Tartu, Estonia
[5]School of Computer Science, University of Birmingham, Birmingham, UK
[6]BRITE Professional Services, Australia

**Abstract**

In the domain of information technology, new paradigms, technologies, development approaches, and application scenarios continue to emerge rapidly presenting new opportunities for researchers to pursue. Emerging paradigms are often considered to be high risk for academic research due to the immaturity of research foundation and ambiguity in fundamentals supporting the paradigm. A pragmatic and holistic strategy would be needed to systematically steer continuous development and evaluation of research in the area and to accelerate impact and adoption. Drawing on our own a decade-long experience in establishing strong and impactful research in cloud computing, we document and reflect on our strategy for advancing research and building a strong community in what was believed to be a high risk emerging/new computing research area for researchers to embark on. The strategy consists of nine key elements which include creation of a roadmap, development of simulators and supporting tools, and creation of application demonstrators that confirm feasibility and effectiveness of applications of the new research findings. We discuss how development of simulators can be cost-effective in accelerating design of real systems and outline strategic role played by different types of publications or conference organization, and development and delivery of educational programs that form part of the research strategy. We illustrate effectiveness of the recommended strategy with our cloud computing research work, which is globally recognized for its impact.

## I. Introduction

Information Technology (IT) is a fertile area for research and new development. It has progressed significantly and gained widespread applications and adoption. Its future potential is vast, and not yet – and never will be - fully known. What new technologies and applications will emerge and which ones will succeed is anybody's guess. Despite its advances, IT still presents several problems and open challenges. Furthermore, several new computing paradigms, technologies and potential high-impact applications scenarios are emerging. These present huge - and yet to be explored - opportunities for researchers and practitioners [1].

To capitalize on the full potential of these new opportunities, researchers need to adopt an effective and holistic strategy for advancing their research in new computing paradigms. But many researchers and research groups fail to visualize a big picture and to generate long-term perspectives; instead, they tend to take a narrow approach, which limit their vision and their success. They also tend to lack adequate drive and required skills for impactful community building for new and emerging computing areas and consequently improve their research quality, productivity, visibility, and transnational impacts [2]. Fostering high quality impactful research calls for strong active communities in the respective field. Strong communities often utilize their "collective research intelligence" to sense pathways and streams that are viable, timely and relevant and to identify and address current and potential future research problems of interest.

In this article, we report on our experience in incubating, developing and systematically evaluating research in emerging paradigm, where we consider cloud computing as a representative case. Our reflection extends more than decade and goes back to 2007, where cloud computing was considered as high risk area for academic research. This is because of the various ambiguities revolving around the technology, lack of rigorous literature describing its foundation and fundamentals, lack of realism, and uncertainty about acceptance and the level of adoption. Emerging areas that are considered to be high risk requires careful systematic navigation into the area to roadmap research, formulate fundamentals, and evaluate the potentials of new directions and to uncover new opportunities for applications and adoptions. A holistic strategy is needed to steer continuous development and evaluation of research, to build a community to accelerate impact and level of adoption.

We share our experience on a strategy that we adopted to advance research and build a prosperous community in emerging and new computing research areas of high risk. In retrospect, the strategy has proven to be effective in accelerating research that matters within the cloud computing area, developing and evaluating sound research foundation, and building an international community of researchers, practitioners and adopters of non-trivial scale. Reflecting on our strategy, we propose nine key elements that accelerated research progression and put cloud computing among the well-established computing areas. The key elements, includes the creation of roadmaps, development of simulators and supporting tools, and developing inspiring application demonstrators that prove feasibility and effectiveness of the new research. Systematic reviews, taxonomies, textbooks and industrial experience reports assist in planning. They can be valuable resource for relevant knowledge, incubators for community building, and steer foundational research, addressing research gaps or inspiring futuristic research into open problems in the field [3]. Such material is often the starting point for researchers to formulate new research problems or to revisit existing ones, as it is often the case for many PhD and MSc theses [4]. To extend current research and to formulate new directions, researchers commonly use highly cited scientific articles or look at industrial problems and experiences [5]. Papers reporting on foundational theories and fundamental research often hold material for empirical investigations, evaluation and/or reflection, attracting interest in adoption [6].

In addition, conferences offer a common platform to bring together interested peers to discuss current research, facilitate networking/connections and accelerate the incubation of new research and communities. Moreover, some international grants facilitate international exchange of researchers paving the way for global visibility and impact. Additionally, collaboration also extended toward authored and edited books, which can be used to teach modules within a curriculum in academia or researchers can use them to understand relevant concepts. Several researchers with different skills set can work together to develop simulators, that are versatile and can be used to study a concept or a technique across various fields, before implementing it in a real environment [7]. Ideas validated through simulations can be put in to practice by incorporating them in software systems that support applications and their deployment in practical settings or production infrastructures.

## II. Key Strategic Elements for Advancing New Computing Areas

We advocate a holistic, pragmatic research strategy that promotes impactful research and community building in new and emerging areas. It would be of interest to researchers in academic institutions, business and public sectors. The strategy, illustrated in Figure 1, comprises of nine key elements, which are discussed in conceptually in this section and illustrated them with practical realisation examples from our cloud computing research in next section.

### 1. Identification of Potential Research Area

The first step is to identify one or more promising new areas for research that the researcher or the research group would like to pursue. Sense the potential of a domain either organically through informal discussions and brainstorms or more methodologically. Methodologically, researchers may, for example, find a potential research domain by reviewing technology trends discussed in formal and informal research forums and those outlined by market-research firms such as Gartner (for example, annual Gartner Hype Cycle for Emerging Technologies) and professional societies such as IEEE Computer Society, and visionary scientific publications in broad domains of interest, national scientific strategies, and perspectives on future research directions. Popular promising new research directions can be considered as a base for community formation. Collaboration with industry practitioners would also help in knowing problems areas that requires further research. Good tutorial articles and industrial experience reports would help get the basic knowledge about the topic. Collective intelligence and vision are the keys for incubating research streams and building research communities that address timely or futuristic problems.

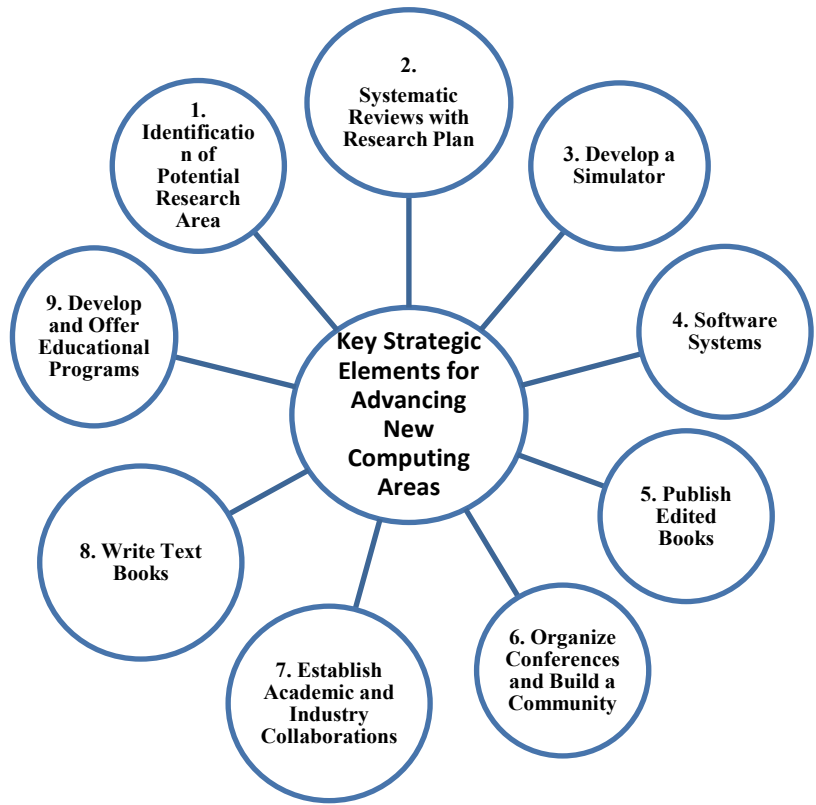

Figure 1: Nine Key Strategic Elements for Advancing Research in New Computing Areas

## 2. Systematic Reviews with Research Plan

Next, it is essential to assess current state of development and potential research problems and challenges in the chosen area. In new/emerging computing areas, systematic reviews help identify the state-of-the-art and knowledge gaps, develop taxonomies, create roadmap and identify future directions. This helps learning about seminal work, starting new research or exploring "blue skies" research. Several established scientific journals value and publish such reviews and feature surveys and/or dedicated special sections. Systematic reviews and assessment play a significant role in accelerating impact, formulating new problems, steering research and acting as a benchmark for comparison. Such publications may sometimes have a detrimental role in steering the research into the wrong direction, in the absence of a responsible peer review system and expert input for scrutinizing the compiled material. The use of Systematic Literature Review (SLR) guidelines [8] helps in ensuring comprehensive coverage and limiting bias in selecting material.

*Research Plan*

Having identified a research area or a topic to pursue, the next crucial and significant activity is to create a research plan and a roadmap. It should include specific areas to be investigated, in the near-term and long-term, timeline, resource requirements and feasibility studies, potential constraints and limitations and anticipated outcomes. Researchers pursue their research work on the chosen area under the guidance of one or more senior researchers and a team leader, closely adhering to the timeline. The research leader takes responsibility to oversee, guide and monitor research progress and research outcomes, which would also include dissemination of research findings to a wider community through publications, seminars, and presentations, and demonstration of application potential of research work and relevance of research. He/she also looks for funding opportunities, formulate specific research proposals and apply for research grants.

## 3. Develop a Simulator

In new/emerging computing areas, researchers are often constrained by the readiness of the technology and availability of required infrastructure to conduct research studies, explore strength and weaknesses, and demonstrate feasibility. In many instances, getting actual physical infrastructure may not be viable due to cost and time constraints and also may not provide all the flexibility that a research study requires. Instead, a simulator is a satisfactory viable option. It can also play a vital role in supporting incubation and growth of research community and to test the proposed methods using lightweight and low-cost environments. Testing new techniques in a real environment is challenging and also costly [16] as one needs invest a lot procure required hardware resources (especially for large-scale experiments) and develop software systems and applications. Researchers use modelling and simulation approach in demonstrating the feasibility of their ideas and carry out experiments to validate their ideas in a controlled environment using simulation tools. Real-world problems can be efficiently tested through simulators as they provide user-friendly environment for exploration and what-if analysis. If off-the-shelf/existing simulator is unable to meet requirements of simulating elements of new computing paradigms, it is advisable to develop a suitable simulator along with graphical user interfaces. A simulator would be immensely helpful to researchers in formulating problems and studying different theoretical models in simulated setups, driving further research and building communities among the respective domain.

4. **Software Systems**

Having evaluated their research, for example, policies/approaches for resource management and application scheduling, through simulation, it is valuable to demonstrate their practical use in real world. This can be achieved by creating appropriate software platforms that would enable potential users in development and testing of real-world applications. These software platforms provide APIs/interfaces for rapid creation of applications and their deployment on a real computing infrastructure. They also support APIs that researchers can use to implement their new resource management and application scheduling polices and evaluate them on real computing systems for real-world applications from domains such as life sciences, health care, and finance.

5. **Publish Edited Books**

As a new research area emerges, several researchers around the world investigate different aspects of it, address challenges associated with that field and publish their findings typically as research papers in conferences or journals. That information, however, remains scattered. To get a quick overview about the emerging area, there is a need for a comprehensive information resource that presents most of the key research work and advances in an integrated manner in a single volume. As no individual researcher will have expertise on all aspects of the emerging field (during its initial days/years), the best option is to edit a new book that brings together contributions from several researchers in the form of individual chapters on different aspects, which are loosely connected. Such edited books serve as initial reference material for researchers, particularly for PhD students and for exploring future directions. Further, edited books will be helpful for academic institutes for introducing relevant modules within their curriculum in degree courses. These books also have been shown to drive interdisciplinary research, by providing case studies, standards and regulations for the respective domains. Edited books also help to bring together researchers in a particular area and create a community that helps in forming collaboration and charting further work. We recommend that senior researchers or the team leader edit a comprehensive book covering many of the related aspects of the research area with contributions from several researchers.

6. **Organize Conferences and Build a Community**

Scientific conferences are very important for dissemination of scientific ideas and act as seeds for high-impact research and building communities. Several studies have shown that physical closeness drives advancements in science and technology and foster research. For example, research on a Paris campus building project [20] showed what occurred when researchers were randomly reassigned to new offices. Researchers began new field-wide partnerships and published groundbreaking articles in high-profile magazines, using a broader selection of keywords relative to prior publications by either partner. Workshops are very common in the research community for qualitative research work embedded within conferences, which provide a platform to learn new techniques, tools and get a chance to participate in brainstorming sessions while solving problems in a group of like-minded people. They also serve as venue for presenting work-in-progress and offer an opportunity for seeking guidance/feedback from research leaders of the area. Further, they can enable industrial and academic collaborations, which can be a foundation for future applied research grants.

Organizations such as IEEE Computer Society and ACM (Association for Computing Machinery) support creation of special interest groups or technical communities that bring together all interested parties to promote new and emerging computing paradigms through establishment of discussion forums, conferences, newsletters, testbeds, guidelines for educational programs, best practices and standards.

7. **Establish Academic and Industry Collaborations**

Academic collaborations drive interdisciplinary research. Additionally, researchers can collaborate with internationally leading researchers especially those who created vision for the field and authored highly cited papers to improve scope and visibility of their research, and to seek opinion and suggestions.

Businesses rely on university researchers for ideas for product developments in new/emerging areas and faculties gain reputation from expanded external research support and collaboration. As much as industry needs new ideas to excel and thrive, scholars require financial and in-kind support from industry to carry out research and maintain professional productivity. These industry collaborations are also becoming mandatory with most of the governmental funding; for example, for Australian Research Council and European Commission research grants research project consortiums are expected to constitute both academic institutions and small and medium-sized enterprises (SMEs). Such collaborations are also needed for facilitating applied research and promoting SMEs and economic growth. Furthermore, industry participation increases the adoptability of the emerging technologies, thus bringing much wider audience/communities onboard.

So, new research teams should try to establish academic and industry collaboration in the chosen area and harness the potential of collaboration.

8. **Write Text Books**

When the field achieves a reasonable maturity, it is timely and critical to develop authored books that present all foundational concepts, technologies, and applications in a consistent and integrated manner. Textbooks provide a good introductory summary or a starting point to a researcher interested to pursue further research. Moreover, they may also provide a consolidated information on previous research about a particular topic as well historical context. Textbooks introduce the new research domains as part of mainstream curriculums thus driving interest in next generation educators and researchers.

9. **Develop and Offer Educational Programs**

It is important to disseminate advances made in a new research and emerging technological developments broadly at Masters and Undergraduate degree programs. This can be achieved by introducing a new subject in existing programs or starting entirely a new program. These programs are also useful to connect someone with the source of information in terms of professional support. For example, a workshop on computational study can provide a chemist an effective platform to learn basic concepts of computational chemistry. Online and offline education programs or workshops can be offered to students and practitioners.

In addition to these strategic elements, researchers need to regularly disseminate their research outcomes broadly using different channels such as trade magazines [13] and social media. By making outcomes of their research work easily accessible to their peers and the broader academic and business communities, they will be able to substantially enhance impact of their work.

**III. Adoption of the Strategy in Our Cloud Computing Research**

During 2005-2008, Grid computing community begun to influence industrial practices towards creation and delivery of computing services as utilities to support enterprise applications. Companies such as Amazon started to offer rentable computing services via its EC2 (Elastic Compute Cloud) initiative. We (Buyya and team) created a SOA (service-oriented architecture)-based container environment that enables the creation of application platforms (such as Aneka) supporting multiple programming models and deployment of their applications on market-oriented cloud computing environments. We established ourselves as a CLOUDS Lab and proposed a reference architecture for market-oriented cloud computing and created software technologies such as CloudSim and Aneka. Around that time other researchers also started sharing their vision for the field.

In this section, we primarily illustrate effectiveness of the recommended research strategy and guidelines which we adopted in our cloud computing research which made significant impact and globally well recognized.

1. *Initial research:* We, like a few other researchers, drew motivation and inspiration to work on cloud computing from the early seminal publications, our article, "Cloud computing and emerging IT platforms: Vision, hype, and reality for delivering computing as the 5th utility," [11] and "A view of cloud computing" [10] from Berkley researchers. These works inspired many further work and emergence of several commercial offerings in cloud computing area. This is evident from more than 10,000 citations received for these and associated family of papers. Thus, this initial research has contributed in building huge communities around cloud computing and relevant research challenges.

2. *Systematic reviews and taxonomies*: Since the emergence of cloud computing field, many key issues such as energy efficiency were investigated and reviewed along with discussion on taxonomies and future directions [21]. With a decade of progress in the field, cloud computing achieved enormous success in industry and business application and opened up new challenges and research directions. In this context, "A manifesto for future generation cloud computing" [12] has also been prepared, by experts from different subfields of cloud computing, which is driving further the community building in the domain, attracting wide interest from young researchers.

3. *Simulations:* We (Buyya and team) have developed a simulator, called CloudSim, for modelling and simulation of cloud computing environments. It helps researchers in evaluation of their new resource management approaches, algorithms, and policies and application scheduling with specific focus on various performance parameters such as security, time, cost, and energy [19]. This simulator also has shown to bring in much wider audience to the cloud computing domain, as jumping into the domain with these simulators is relatively easy as illustrated by its use in over 5000 research papers published by researchers worldwide.

4. *Software systems:* Researchers are interested to move their work "bench to market" - to transition their work in to a product or a real-world application. Prominent software systems within the cloud computing domain such as Aneka, OpenStack, and Hadoop have driven the research in the respective domains significantly and helped in bringing people from other communities such as data analytics, distributed computing and scientific computing, to the cloud domain. In Cloud computing domain, new resource management and application scheduling algorithms/policies we originally created and evaluated using CloudSim [22] are demonstrated/implemented for practical deployment by plugging them to real software systems such as Aneka and OpenStack [23].

5. *Edited books:* The prominent researchers within the domain of cloud computing are editing books time to time, which can help researchers to generally understand the concepts. Edited books such as "Cloud Computing: Principles and Paradigms" published in the early days of the field helped researchers to understand the basic concepts and initial research, and are being used as reference material in several courses and curriculums. Later with further advances in the field, books such as "*Encyclopedia of Cloud Computing [18]*" complemented it.

6. *Conferences and Community:* In its early days several new conferences dedicated to Cloud computing such as IEEE/ACM UCC, CCGrid, IEEE Cloud and ACM SoCC were established. IEEE also established community forums such as the IEEE Technical Committee on Cloud Computing. These provided opportunity to academics, researchers, practitioners to discuss ongoing research and helped new researchers to discuss their ideas with the invited speakers and co-participants to find a chance to collaborate with them.

7) *Academic and Industrial Collaborations:* Educational institutions are collaborating and submitting collaborative research grant proposals both nationally and globally. For instance, European Union has funded a number of collaborative research projects in the cloud computing domain during the past decade [9]. In addition to facilitating research collaboration (both academic and industry), these projects help participating students to visit and work with other research groups which enables them to discuss and improve their work. Moreover, the SMEs and startups resulting from these projects have been the early adapters of cloud computing technology and applications, and drove local economies and Cloud adaption policies at local governments. We collaborated with companies such as CA and Samsung and developed a new framework for ranking of Cloud services, which the Cloud Service Measurement Index Consortium (CSMIC) adopted it.

8) *Text books*: Like few others, we wrote a text book "Mastering Cloud Computing", which introduces foundations, technologies, programming examples, and applications, in a much more unified and consistent manner for broader audience including. It is adopted as a text for Under Graduate (UG) and Post Graduate (PG) courses in several universities worldwide.

9) *Educational programs*: Several educational programs and online courses are offered to deliver the required cloud computing knowledge to the researchers. Various renowned universities are working together to run these education programs to teach basic and advanced and emerging trends. In the University of Melbourne, a dedicated Masters' level program in Distributed Computing was established which now educates close to 500 students each year.

All these demonstrates how the recommended strategic elements have collectively driven our cloud computing research journey from infancy to a well-established one that made tremendous impact in academic and business world.

### IV. Summary and Conclusions

Drawing on our successful experience, we shared a holistic strategy for advancing research and its impact in new and emerging computing paradigms. We have discussed the role of supporting infrastructure that we developed and used - simulators and software systems, and the role of various types of publications including systematic reviews, taxonomies, research articles and books in supporting community building and accelerating the incubation of research. It is critical to establish community forums and develop educational programs. We illustrated application of our strategy to Cloud computing field from its emerging days (in 2008) to till date. We believe the proposed strategy would serve as a catalyst for new research initiatives to accelerate their research and its impact with confidence.

**ABOUT THE AUTHORS**

**RAJKUMAR BUYYA** is a Redmond Barry Distinguished Professor, Director of the Cloud Computing and Distributed Systems (CLOUDS) Laboratory at the University of Melbourne, Australia and CEO of Manjrasoft Pty Ltd. He is recognized as a "Web of Science Highly Cited Researcher" for six consecutive years since 2016, a Fellow of IEEE and Scopus Researcher of the Year 2017 with Excellence in Innovative Research Award by Elsevier. He has been recognised as the "Best of the World" twice for research fields (in Computing Systems in 2019 and Software Systems in 2021) as well as "Lifetime Achiever" and "Superstar of Research" in "Engineering and Computer Science" discipline twice (2019 and 2021) by the Australian Research Review. He is serving as Co-Editor-in-Chief of Journal of Software: Practice and Experience established 50+ years ago. Contact him at rbuyya@unimelb.edu.au.

**SUKHPAL SINGH GILL** is a Lecturer (Assistant Professor) in Cloud Computing at School of Electronic Engineering and Computer Science, Queen Mary University of London, UK. His research interests include Cloud Computing, Fog/edge Computing and Energy Efficiency. He received a Doctor of Philosophy (PhD) in Computer Science from Thapar Institute of Engineering and Technology, India. He is a professional member of ACM. Contact him at s.s.gill@qmul.ac.uk

**SATISH NARAYANA SRIRAMA** is an Associate Professor at School of Computer and Information Sciences, University of Hyderabad, India and a Visiting Professor and the honorary head of the Mobile & Cloud Lab at the Institute of Computer Science, University of Tartu, Estonia, which he led as a Research Professor until June 2020. His current research focuses on cloud computing, mobile cloud, and migrating scientific computing and large scale data analytics to the cloud. He received his Ph.D. in computer science from RWTH Aachen University in 2008. He is an IEEE Senior Member. Contact him at satish.srirama@uohyd.ac.in.

**RAMI BAHSOON** is a Reader at the School of Computer Science, University of Birmingham, UK. He received a Doctor of Philosophy (PhD) in Software Engineering from University College London in 2006. His research interest includes software architecture, cloud and services software engineering. He is a fellow of the Royal Society of Arts and Associate Editor of IEEE Software - Software Economies. Contact him at r.bahsoon@cs.bham.ac.uk.

**SAN MURUGESAN** is the Director of BRITE Professional Services; an Adjunct Professor at Western Sydney University, Australia; and former editor-in-chief for IEEE Computer Society's IT Professional magazine. He is a Senior Consultant with Cutter Consortium in USA, a former Senior Research Fellow of the US National Research Council at NASA Ames Research Center in California, and a consultant, corporate trainer, researcher, academic and author. Contact him at san1@internode.net.